\documentclass[11pt,sort&compress]{elsarticle}

\makeatletter
  \long\def\pprintMaketitle{\clearpage
  \iflongmktitle\if@twocolumn\let\columnwidth=\textwidth\fi\fi
  \resetTitleCounters
  \def\baselinestretch{1}%
  \printFirstPageNotes
  \begin{center}%
 \thispagestyle{pprintTitle}%
   \def\baselinestretch{1}%
    {\large\bf\@title}\par\vskip5pt
    \normalsize\elsauthors\par\vskip5pt
    \footnotesize\itshape\elsaddress\par\vskip10pt
    \end{center}%
  \gdef\thefootnote{\arabic{footnote}}%
  }
\makeatother

\newcommand\blfootnote[1]{%
  \begingroup
  \renewcommand\thefootnote{}\footnote{#1}%
  \addtocounter{footnote}{-1}%
  \endgroup
}

\journal{}
\usepackage{amsmath}
\usepackage{amsfonts}
\usepackage{lipsum}
\usepackage{graphicx}
\usepackage{xspace}
\usepackage{epsf}
\usepackage{setspace}
\usepackage{float}
\usepackage{abstract}
\usepackage{booktabs}
\usepackage[]{siunitx}
\usepackage{longtable}
\usepackage{placeins}

\usepackage[]{microtype}

\usepackage[margin=1in]{geometry}

\usepackage{stmaryrd}
\SetSymbolFont{stmry}{bold}{U}{stmry}{m}{n}

\usepackage[dvipsnames]{xcolor}
\usepackage{bm}

\newcommand{\transpose}{\mathsf{T}}
\definecolor{lightblue}{rgb}{0.63, 0.74, 0.78}
\definecolor{seagreen}{rgb}{0.18, 0.42, 0.41}
\definecolor{orange}{rgb}{0.85, 0.55, 0.13}
\definecolor{silver}{rgb}{0.69, 0.67, 0.66}
\definecolor{rust}{rgb}{0.72, 0.26, 0.06}
\definecolor{purp}{RGB}{68, 14, 156}

\colorlet{lightrust}{rust!50!white}
\colorlet{lightorange}{orange!25!white}
\colorlet{lightlightblue}{lightblue}
\colorlet{lightsilver}{silver!30!white}
\colorlet{darkorange}{orange!75!black}
\colorlet{darksilver}{silver!65!black}
\colorlet{darklightblue}{lightblue!65!black}
\colorlet{darkrust}{rust!85!black}
\colorlet{darkseagreen}{seagreen!85!black}

\usepackage[export]{adjustbox} 

\usepackage{hyperref}
\hypersetup{
  colorlinks=true,
}

\usepackage[labelfont=bf,font=small]{caption}

\usepackage[nameinlink]{cleveref}
\crefname{equation}{}{}

\usepackage[parfill]{parskip}

\begin{document}

\hypersetup{
  linkcolor=darkrust,
  citecolor=seagreen,
  urlcolor=darkrust,
  pdfauthor=author,
}

\begin{frontmatter}

\title{{\large\bfseries Transmission of High-Amplitude Sound through Leakages of Ill-fitting Earplugs}}

\author[1,2]{Haocheng Yu}
\ead{haochey@gatech.edu}
\author[1,3]{Krishan K.\ Ahuja}
\author[1]{Lakshmi L.\ Sankar}
\author[1,2,4]{Spencer H.\ Bryngelson}

\address[1]{Daniel Guggenheim School of Aerospace Engineering, Georgia Institute of Technology, Atlanta, GA 30332, USA\vspace{-0.15cm}}
\address[2]{School of Computational Science \& Engineering, Georgia Institute of Technology, Atlanta, GA 30332, USA\vspace{-0.15cm}}
\address[3]{Georgia Tech Research Institute, Atlanta, GA 30332, USA\vspace{-0.15cm}}
\address[4]{George W. Woodruff School of Mechanical Engineering, Georgia Institute of Technology, Atlanta, GA 30332, USA}

\date{}
\end{frontmatter}
\blfootnote{
Code available at: \url{https://github.com/MFlowCode/MFC}
}

\begin{abstract}

High sound pressure levels (SPL) pose notable risks in loud environments, particularly due to noise-induced hearing loss.
Ill-fitting earplugs often lead to sound leakage, a phenomenon this study seeks to investigate. 
To validate our methodology, we first obtained computational and experimental acoustic transmission data for stand-alone slit resonators and orifices, for which extensive published data are readily available for comparison.
We then examined the frequency-dependent acoustic power absorption coefficient and transmission loss (TL) across various leakage geometries, modeled using different orifice diameters. 
Experimental approaches spanned a frequency range of \SIrange[]{1}{5}{\kilo \hertz} under SPL conditions of \SIrange[]{120}{150}{\decibel}. 
Key findings reveal that unsealed silicone rubber earplugs demonstrate an average TL reduction of approximately \SI{18}{\decibel} at an overall incident SPL (OISPL) of \SI{120}{\decibel}. 
Direct numerical simulations further highlight SPL-dependent acoustic dissipation mechanisms, showing the conversion of acoustic energy into vorticity in ill-fitting earplug models at an OISPL of \SI{150}{\decibel}.
These results highlight the role of earplug design for high-sound-pressure-level environments.
\end{abstract}

\section{Introduction}\label{s:intro}

To improve the performance of earplugs, understanding the acoustics of sound leakage through any air gaps in ill-fitting earplugs is necessary.
These air gaps serve as pathways for sound to travel.
\Citet{cummings1991impedance} investigated the effects of air gaps between impedance tube test samples and their sample holders to determine the accuracy of impedance tube measurements of porous sound-absorbing materials.
This work exposed the importance of leakage around impedance tube sample holders on the acoustic properties of the test samples.
The author provided evidence that small leaks in impedance tubes notably increase the sound absorption of the test sample, presumably through increased transmission around the test sample and the sample holder. 
This work focused on the influence of leakage and measurements of flow resistivity, a material property of sound-absorbing materials.
This work did not quantify the transmission loss (TL) of leaks nor the influence of varying sizes and geometries of leak paths, which is studied in the present paper.

Many earplugs are made from sound-absorbing materials (for example, porous foam) because they are soft enough to fit comfortably inside ear canals.
Three other commonly used earplugs are available: wax, universal-fit flanged types, and custom-molded. 
The universal-fit flanged type is typically made from non-absorbing materials, such as silicone rubber, which has a relatively high stiffness.
The surface of such an earplug acts like an acoustically rigid boundary whose impedance approaches infinity. 
Limited studies addressing leak paths around these earplugs have been published. 
However, this lack of research does not imply that leak paths are unimportant to the performance of this type of earplug.
For example, \citet{kvaloy2010comparison} found weaker attenuation in silicone earplugs compared to foam earplugs. 
\citet{alt2012objective} reported the leak dimensions among different types of earplugs by measuring the volumetric flow rate of water through various leak paths. 
Their results show that custom-molded earplugs permit 50\% lower volumetric water flow through flanking paths than universal-fit earplugs. 
\Citet{tufts2013attenuation, tufts2013attenuationconf} published a study regarding the comfort and effectiveness of custom-molded earplugs. 
They showed that custom-molded earplugs must be inserted deeply into the second bend of the ear canal to achieve \SI{20}{\decibel} more attenuation; however, all of their experimental users reported discomfort. 
Even though leak paths can be decreased by customizing and extending the plugs into ear canals, \citet{harvie2024perforations} reported that user comfort was compromised, and long-term use could lead to tympanic membrane perforation. 
Besides their flexibility, the most elementary difference between a stiff rigid earplug and a sound-absorbing one is that no acoustic transmission ($\mathrm{TL} \ge \SI{60}{\decibel}$) is viable through the rigid earplugs, according to analytical models.

In high sound pressure level (SPL) environments, where nonlinear acoustic phenomena may occur, a dominant dissipation mechanism identified in resonant acoustic liners is vortex shedding, which occurs at the mouths of the resonators when high-SPL sound waves propagate through them~\cite{tam2001numerical, leung2007duct, tam2008numerical, salikuddin1986nonlinear, dai2012discrete}.
Artificial flanking paths, or orifices, are often introduced purposely to enhance an earplug's TL at higher SPL.
At the same time, they increase transparency in a low-SPL environment.
\citet{berger2008empirical} used an impulse method to compare the TL of different earplugs with small openings being introduced. 
A TL up to \SI{25}{\decibel} with a pulse peak of \SI{190}{\decibel} was detected, while trivial attenuation was observed with a pulse peak of \SI{110}{\decibel}.
These earplugs with orifices have been fielded for use by military users.
Understanding and decreasing the effects of leak paths on rigid earplugs is beneficial for reducing the risk of temporary or permanent noise-induced hearing loss in high-SPL environments, where earplugs made from non-absorbing materials may be used (for example, \citet{abel2004sound}).

This study aims to understand sound absorption within small leaks under high incident sound pressure levels (ISPL).
\Cref{s:background} reviews prior literature on modeling and testing leakage in earplugs while introducing an axisymmetric model for earplug leakage.
\Cref{s:methods} outlines the experimental and computational methodologies employed in this study.
The experimental and computational results are analyzed in \cref{s:results}, followed by concluding remarks in \cref{s:conclusions}.

\section{Background}\label{s:background}

\subsection{Previous Work}\label{ss:previous_work}
In many studies, circuit-analogy models are used to represent the damping and force response phenomenon when sound propagates through nonuniform and discontinuous systems such as the middle ear transmission~\cite{rosowski1996models, o2008middle, ayat2013using}.
\citet{henriksen2008using} published a related study on earplug leakage using a model analogous to a circuit with the impedance of the leakage and the earplug in parallel. 
This model is based on Rschevkin's acoustic impedance model~\cite{rschevkin1964course} for a circular orifice with non-normalized scales.
He measured the attenuation of the earplugs using a two-microphone (2-mic) system with one microphone on the incident side and another on the transmitted side. 
Henriksen's model agreed with his experimental results, with a mean difference of \SI{-0.7}{\decibel}, indicating that acoustic attenuation increased with increasing frequency.
However, the attenuation was reported only at low frequencies (\SIrange[]{50}{1000}{\hertz}), which do not primarily encompass the most sensitive audible frequency band of human hearing. 
Earplugs and other types of filling objects can have their attenuation measured without knowing the size of the leaks; however, \citet{henriksen2008using} found that small differences in leak size notably diminish the effectiveness of hearing protection.

\Citet{groon2015air} aimed to understand leakage effects on the acoustic measurement inside realistic human ears and collected data from live human ear canals using otoacoustic emission, which has been widely used in clinical settings to detect the acoustic response of the cochlea~\citep{kemp1990guide}.
However, quantitative studies did not focus on leaks introduced by earplugs or high SPL conditions.
\citet{yu2024numerical} computationally studied the earplug leakage effects in high SPL conditions by modeling the earplug as a fluid with a density similar to that of liquid water.
While they observed greater acoustic attenuation under higher SPL conditions, no evidence of vortex shedding around the leak paths was found. 
These studies primarily focused on the leakage of the measuring apparatus and its impact on measurement accuracy inside human ears. 
Our study fills these knowledge gaps by systematically developing an analytical approach and providing quantitative insights through both experimental and numerical methods. 
The analysis spans a wider acoustic frequency range (\SIrange[]{1}{6}{\kilo \hertz}), including the most sensitive human hearing range (\SIrange[]{2}{5}{\kilo \hertz}) \cite{flood2010essentials}.

\subsection{Circular Orifice Model of Leaks}\label{ss:analy_model}

\begin{figure}[htb]
    \centering
    \includegraphics[]{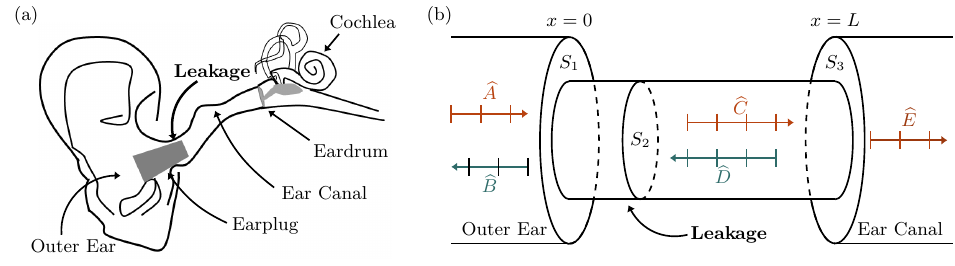}
    \caption{
        Sketch of (a) the human ear with an earplug inserted in the ear canal and (b) its analogy to a circular orifice opening
    }
    \label{LeakSchematics}
\end{figure}

The leaks around the earplugs in \cref{LeakSchematics} are similar to small openings, such as variable-size orifices between two isolated rooms.
The acoustic transmission and absorption characteristics of small orifices have been widely studied, highlighting that acoustic performance depends on orifice geometry, including size, length, and cross-sectional shape~\cite{ingaard1950acoustic, cummings1984acoustic, leung2007duct}.
\Citet{gaeta2016effect} further demonstrated substantial variability in acoustic power absorption due to subtle geometric differences among orifices.
However, these studies focused on well-defined orifice shapes, and a knowledge gap remains for irregular leak geometries that represent faithful earplug scenarios.
Due to the absence of a representative leak shape, we performed theoretical predictions by drawing an analogy between a leak path and a circular orifice opening without considering the existence of the eardrum to quantify the effects of leaks around a given earplug.
We develop a purely analytical model of the acoustic transmission through orifice openings. The model is based on the transfer matrix model (TMM) for its computational efficiency and flexibility in handling complex acoustic boundary conditions \cite{lee2009modified, brouard1995general}.
This model aims to mimic leaks with an anechoic termination, which isolates the orifice opening from other reacting surfaces.

All acoustic waves are assumed to be plane waves throughout this study, as the transverse dimensions of the ear canal and leaks are small enough to support plane wave propagation in the frequency domain of interest.
The acoustic wave interactions at both interfaces ($x = 0$ and $x = L$) around the orifice leak path (illustrated in \cref{LeakSchematics}) are governed by pressure and volumetric flow continuity conditions.
The leak path length $L_\mathrm{eff} = L + 0.821 r$ is corrected with end correction for flanged circular pipes based on \citet{kinsler2000fundamentals}, where $r$ is the radius of the orifice openings for mass loading effects.
Under the assumption of plane-wave propagation, the boundary conditions form a coupled linear system
\begin{gather}
    \mathbf{M} \, \mathbf{\widehat{p}} = \mathbf{0},
    \label{eq:conservation}
\end{gather}
where vector $\mathbf{\widehat{p}} = [\widehat{A}, \widehat{B}, \widehat{C}, \widehat{D}, \widehat{E}]^\transpose$
contains the acoustic wave amplitude terms defined in \cref{LeakSchematics}, and $\mathbf{M}$ is the coupling matrix
\begin{align}
    \mathbf{M} &= \begin{bmatrix}
      1 & 1 & -1 & -1 & 0 \\[4pt]
      S_{1} & -S_{1} & -S_{2} &  S_{2} & 0 \\[4pt]
      0 & 0 & e^{-ikL_{\mathrm{eff}}} & e^{ikL_{\mathrm{eff}}} & -e^{-ikL_{\mathrm{eff}}} \\[4pt]
      0 & 0 & S_{2}e^{-ikL_{\mathrm{eff}}} & -S_{2}e^{ikL_{\mathrm{eff}}} & -S_{3}e^{-ikL_{\mathrm{eff}}}
    \end{bmatrix},
    \label{eq:couple_matrix}
\end{align}
where $S_1$, $S_2$, and $S_3$ are the cross-sectional areas of the outer ear, leakage, and ear canal as shown in \cref{LeakSchematics}.
The first and second rows of $\mathbf{M}$ in \cref{eq:couple_matrix} are the pressure and volumetric flow conservation equations at the interface of $x = 0$ seen in \cref{LeakSchematics}. 
The third and fourth rows of $\mathbf{M}$ in \cref{eq:couple_matrix} are the conservation equations at the interface $x = L$.
The generalized acoustic wavenumber $k$ depends on the viscous model and the shape of the leak path.
Assuming that the leak path is circular, this study uses Blackstock's viscous model~\cite{blackstock2001fundamentals}:
\begin{equation}
    k = \frac{\omega}{c_0}\left[1+\frac{(1-i)}{r}\sqrt {\frac{\mu}{2 \rho_0 \omega}} \left(1 + \frac{\gamma - 1}{\sqrt{\mathrm{Pr}}}\right)\right],
    \label{blackstock}
\end{equation} 
where $\omega = 2 \pi/f$ is the angular frequency, $i$ is the imaginary unit, $c_0$ is the speed of sound, $\mu$ is the dynamic viscosity, $\rho_0$ is the ambient density, $\gamma$ is the specific heat ratio, and $\mathrm{Pr}$ is the Prandtl number.
The system of equations \cref{eq:conservation} can be solved for the pressure transmission coefficient $|\widehat{E}/\widehat{A}|^2$, which is the ratio of the transmitted to the incident power intensity and will be used to calculate the TL.
As no literature was found that measures and generalizes the shape of leak paths around earplugs, our models for leak paths are not claimed to represent what may exist around real earplugs generally. 
The model simplifies leak path geometries and assumes idealized boundary conditions, which may limit the direct application to irregular leak geometries found in practice.

\section{Methodology}\label{s:methods}

\subsection{Acoustic Performance Metrics}\label{ss:metrics}

The following discussion applies only to plane-wave, normal-incidence impedance tubes, such as the one described in this work.
Here, two metrics of acoustic performance were used: the sound power absorption coefficient and transmission loss (TL).
The sound power absorption coefficient is defined as
\begin{equation}
    \alpha = 1 - |\widehat{R}|^2 - |\widehat{T}|^2,
    \label{eq:alpha}
\end{equation}
where $|\widehat{R}|^2$ and $|\widehat{T}|^2$ are the power reflection and transmission coefficients, and 
\begin{equation}
    \mathrm{TL} = 10 \log_{10}\frac{1}{\tau},
    \label{eq:TL_tau}
\end{equation}
where $\tau$ is the ratio of the transmitted to the incident power spectral density (PSD), which is the SPL spectra of the sound.
The power absorption coefficient measures the dissipation at the specific sound level of a source as it passes through an acoustic leakage. 
In contrast, the TL of sound measures the reduction in sound level of a sound source as it passes through an acoustic leakage. 
TL is the number of decibels stopped by the acoustical barrier, such as a leak or a wall, and is measured at different frequencies.

Regarding the sound source, the incident sound pressure level (ISPL) represents the decibel sound level at a specific frequency and is defined as 
\begin{equation}
    \mathrm{ISPL} = 20 \log_{10}\left(\widehat{p}/\widehat{p}_\mathrm{ref}\right),
    \label{eq:ISPL}
\end{equation}
where $\widehat{p}$ is the root-mean-squared pressure (RMS) of the sound, and $\widehat{p}_\mathrm{ref} = \SI{20}{\micro \pascal}$ is the standard reference pressure in air.
The overall incident pressure level (OISPL) represents the decibel sound level across broadband frequencies and is obtained from
\begin{equation}
\mathrm{OISPL} = 10\log_{10}\!\left(\frac{\sum_{i=1}^{N}\widehat{p}_i^{\,2}}{\widehat{p}_{\mathrm{ref}}^{\,2}}\right),
    \label{eq:OISPL}
\end{equation}
where $\widehat{p_i}$ is the RMS pressure of the sound at the $i$-th frequency band, and $N$ is the total number of frequency bands considered.
In this study, we use the OISPL to specify the nominal sound level of impulsive and broadband sources and the ISPL to represent the sound level of discrete tones at a specific frequency. 

\subsection{Experimental Method} \label{ss:exp_method}

\subsubsection{Experimental setup} \label{sss:exp_setup}

\begin{figure}[htp]
    \centering
    \includegraphics[]{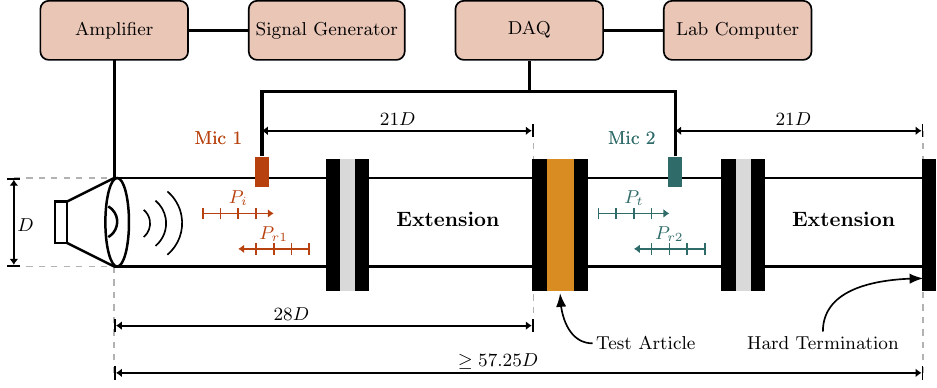}
    \caption{
        A layout (not to scale) and specifications of the extended two-sided tube with its data acquisition system.
        The inner diameter of the tube is $D = \SI{25.4}{\milli\meter}$.
    }
    \label{Exp_tube_schematics} 
\end{figure}

We employed a 2-mic impulsive method as described by \citet{yu2022modelling} to measure the TL of the modeled earplug leakage.
This method is similar to that used by \citet{salikuddin1980impulse} and is validated extensively by, for instance, \citet{salikuddin1979impulse,salikuddin1980impulse} and \citet{davies1979impulse}.
These investigations showed good agreement between impulse-technique measurements and wave-decomposition results. 
They demonstrated that an impulsive sound source and a single microphone are sufficient to construct a two-sided tube.
The measurements were conducted using a two-sided impedance tube with a single microphone on each side.
A sketch of the test setup appears in \cref{Exp_tube_schematics}, and a brief description of the setup is provided below.
The inner diameter ($D$) of the impedance tube is \SI{25.4}{\milli\meter}.
The tube is smooth and long enough for evanescent higher-order modes to decay.
The first higher-order mode cuts off at \SI{7.92}{\kilo\hertz}; all measurements ($\leq$\SI{5}{\kilo\hertz}) are safely within the plane-wave regime.

Acoustic waves were recorded using PCB 378A12 microphones at a sampling rate of \SI{51200}{\hertz}. 
Microphone sensitivities (in \si{\milli\volt\per\pascal}) were determined using a Larson Davis CAL200 precision acoustic calibrator at a frequency of \SI{1}{\kilo\hertz} and a sound pressure level (SPL) of \SI{94}{\decibel}.
The manufacturer-rated maximum limit for the 378A12 microphones is \SI{173}{\decibel}, which exceeds the highest measured peaks of \SI{168}{\decibel} in this study. 
Thus, all recordings remained within the linear dynamic range of the sensors, and no clipping was observed.

\begin{figure}[ht]
    \centering
    \includegraphics[]{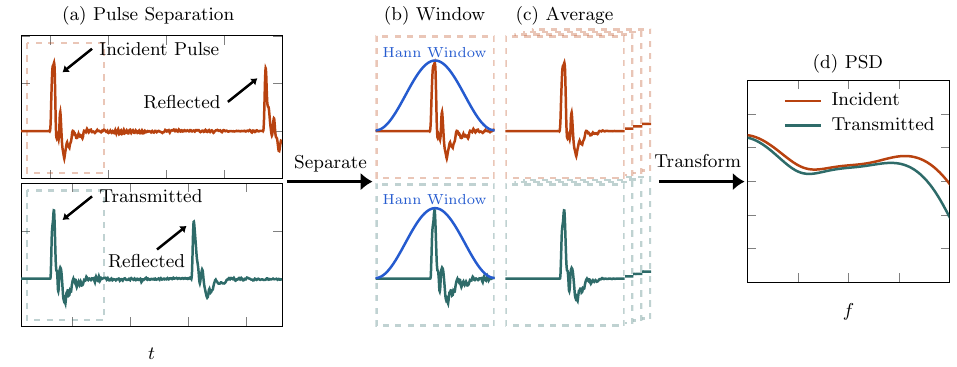}
    \caption{
        The processing steps: (a) separating the incident and transmitted pulse from their reflected pulses, (b) applying a Hann window on the separated incident and transmitted pulses, (c) averaging 100 windowed incident and reflected pulses, and (d) using FFT to determine the incident and transmitted power spectral density (PSD), which is used to calculate the TL defined in \cref{eq:TL_tau}.
    }
    \label{Exp_pulse_separation} 
\end{figure}

\Cref{Exp_pulse_separation} illustrates the signal processing of the measured pulses. 
We separated the incident time-domain zero-padded pulse signals from their reflected pulses at each microphone.
One hundred separated pulses were Hann-windowed and averaged. 
Spectral analysis was performed with the FFT method~\cite{brigham1988fast}, using 1,000-point Hann-windowed blocks with 50\% overlap, resulting in a frequency resolution of $\Delta f = \SI{51.2}{\hertz}$.
The nominal overall incident sound pressure level (OISPL) for this study is 120, 130, 140, and \SI{150}{\decibel}.
\Cref{table:measured_SPL} shows the peak SPL and OISPL of the impulsive signals for three nominal test conditions.

\begin{table}[htb]
    \centering
    \caption{
        Measured incident sound pressure level.
        \label{table:measured_SPL}
    }
    \begin{tabular}{c c c}
        \toprule
        Nominal  & Measured  & Measured \\
        OISPL (\si{\decibel}) & Peak SPL (\si{\decibel}) & OISPL (\si{\decibel})\\
        \midrule
        120 & 140.0 & 121.7 \\ 
        130 & 147.5 & 130.7 \\ 
        140 & 159.9 & 139.2 \\ 
        150 & 167.9 & 147.8 \\
        \bottomrule
    \end{tabular}
\end{table}

\subsubsection{Transmission Loss of Orifice Plates}\label{sss:exp_orifice}

\begin{figure}[ht]
    \centering
    \includegraphics[]{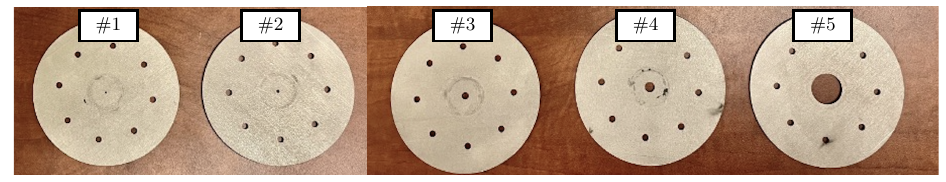}
    \caption{The five orifice plates studied in this work.
    }
    \label{Exp_orifice_photo} 
\end{figure}

We conducted measurements of transmission loss (TL) in thin steel plates with circular orifices (denoted by Test Articles \#1--5 in \cref{Exp_orifice_photo}) at an overall incident SPL (OISPL) in \SI{120}{\decibel} to verify our experimental method against our analytical model described in \cref{ss:analy_model}.
The specifications of the tested orifice plates are tabulated in \cref{table:TestArticle}.

\begin{table}[htb]
    \centering
    \caption{
        Specifications of the steel orifice plates studied in this work. We use steel as the raw material to machine all solid and orifice plates, ensuring a flush surface finish.
        All the orifice plates have an outer diameter of $4D$ and a thickness of $0.125 D$.
        \label{table:TestArticle}
    }
    \begin{tabular}{ l r r r}
        \toprule
        \# & Test Article & Inner Diameter & Area Ratio \\ 
        \midrule
        \#1 & 0.5\% Orifice Plate & $0.071 D$ & 0.005 \\ 
        \#2 & 1\% Orifice Plate &  $0.100 D$ & 0.010 \\ 
        \#3 & 5\% Orifice Plate & $0.224 D$ & 0.050 \\ 
        \#4 & 10\% Orifice Plate & $0.316 D$ & 0.100 \\ 
        \#5 & 100\% Orifice Plate &  $D$ & 1.000 \\
        \bottomrule
    \end{tabular}
\end{table}

\begin{figure}[htp!]
    \centering
    \includegraphics[]{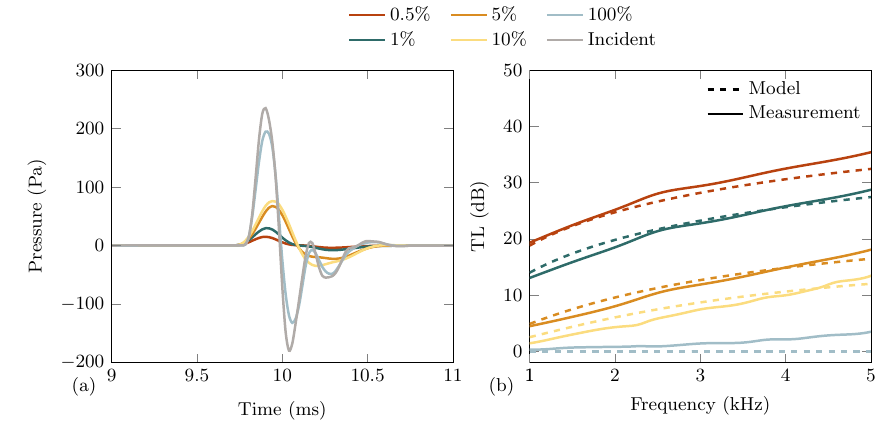}
    \caption{Measured performance of the steel orifice plates (test article \#1--5) under $\mathrm{OISPL}=\SI{120}{\decibel}$ for (a) the transmitted pulse in time domain and (b) the transmission loss spectrum compared to the analytical model described in \cref{ss:analy_model}.
    }
    \label{Exp_OrificeTest} 
\end{figure}

\Cref{Exp_OrificeTest} shows the time-domain signals and the TL spectra for plates with orifice area ratios of 0.5\%, 1\%, 5\%, 10\%, and 100\%.

The measured TL trends generally align well with analytical predictions.
However, discrepancies of approximately \SIrange[range-units = single]{3}{4}{\decibel} are observed at higher frequencies. 
This deviation is primarily attributed to elevated viscous losses within the impedance tube system at higher frequencies, a factor not accounted for in the simplified analytical model.

\subsubsection{Ear Test Articles}\label{sss:exp_test_article}

\begin{figure}[ht]
    \centering
    \includegraphics[]{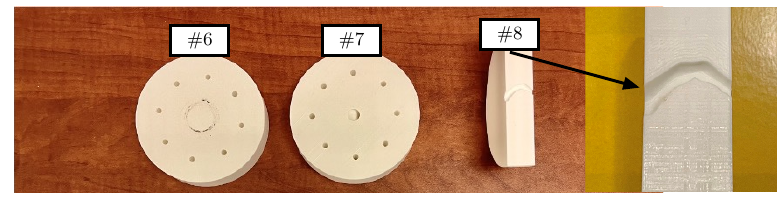}
    \caption{The three test articles evaluated in this work, with a magnified view of Test Article \#8 showing the modeled ear canal geometry.}
    \label{TestArticles} 
\end{figure}

We 3D printed several ear test articles to investigate the effects of leakage on transmission loss in a modeled ear canal.
\Cref{TestArticles} shows the 3D-printed test articles. 
Test Article \#6 is a 3D-printed solid with a thickness of \SI{24.38}{\milli\metre}, serving as a baseline for comparison with the modeled ear canal.
Test Article \#7 represented a 3D-printed modeled ear canal with a circular cross-sectional shape derived from the work of \citet{stinson1989specification}, which quantified the cross-sectional area of human ear canals as a function of axial position. 
It should be noted that the 3D-printed ear canal geometry excludes the anatomic characteristics, such as elasticity and surface roughness, present in real human ear canals.
Test Article \#8 is a halved version of the modeled ear canal, designed to visualize and examine the geometry of a prototypical human ear canal. 
The details and implications of its design will be discussed further in \cref{s:results}.

\subsection{Numerical Methods}\label{ss:dns}

\subsubsection{Compressible Flow Solver}\label{sss:sim_method}

To link the effects of the leakage and conversion into vorticity as a dissipation mechanism, we use direct numerical simulation (DNS) based on the open-source code MFC~\cite{bryngelson2021mfc, wilfong2025mfc}.
MFC solves governing partial differential equations for compressible multi-fluid multi-phase flows.
We use the compressible Navier--Stokes equations, as presented in \cref{eq1,eq2}, as the governing equations for this study.
It consists of mass, momentum, and energy conservation for the fluid, as well as a gas equation of state.
This model can capture the propagation of sound waves through small air leaks.
The governing equations we use are
\begin{gather}
    \frac{\partial \mathbf{q}}{\partial t} + \nabla \cdot \mathbf{F}(\mathbf{q}) = \mathbf{s}(\mathbf{q}) 
    \label{eq1}
\end{gather}
where $\mathbf{q}$ is the state variable vector, $\mathbf{F}$ is the flux tensor, and $\mathbf{s}$ are the source terms.
The representation of the variables of the PDE is 
\begin{align}
    \mathbf{q} &= \begin{bmatrix}
        \rho \\
        \rho \mathbf{u} \\
        \rho E \\
    \end{bmatrix}, \quad
    \mathbf{F} = \begin{bmatrix}
        \rho \mathbf{u} \\
        \rho \mathbf{u} \mathbf{u}   + p\mathbf{I} - \mathbf{T} \\
        (\rho E + p) \mathbf{u} - \mathbf{T} \cdot \mathbf{u} \\
    \end{bmatrix}, \quad 
    \mathbf{s} = \begin{bmatrix}
        f_g f_{\delta}/a \\
        f_g f_{\delta} \mathbf{r} \\
        0 \\
    \end{bmatrix}
    \label{eq2}
\end{align}
where $\rho$, $\mathbf{u}$, and $p$ are the density, velocity, and pressure, $E = e + \|\mathbf{u}\|^2/2$ is the total energy of the fluid, $e$ is the internal energy, $\mathbf{I}$ is the identity matrix, and $\mathbf{T}$ is the viscous stress tensor with negligible bulk stresses
\begin{gather}
    \mathbf{T} =\frac{1}{\mathrm{Re}} \left( \nabla \mathbf{u} + (\nabla \mathbf{u})^{\transpose} -\frac{2}{3}(\nabla \cdot \mathbf{u})\mathbf{I} \right),
    \label{eq3}
\end{gather} 
where $\mathrm{Re} = c_0 d/\nu$ is the Reynolds number based on the speed of sound $c_0$, the length of the leak path $d$ as the characteristic length to scale the simulation domain, and the dynamic viscosity of air $\nu$.
The nonconservative source terms in $\mathbf{s}$ corresponding to the one-way plane wave source of \citet{maeda2017source}, where $f_g$ is the time-dependent amplitude for monopole and dipole sources, $f_{\delta}$ is the Gaussian monopole support function, $a$ is the speed of sound, and $\mathbf{r}$ is the unit vector.
The equation system is closed by a gas equation of state, which is $p = \left(\gamma - 1\right) \rho e$.
For this study, all sound waves are assumed to be harmonic, meaning that the time-dependent amplitude $f_g$ is sinusoidal with time.

\begin{figure}[htb]
    \centering
    \includegraphics[]{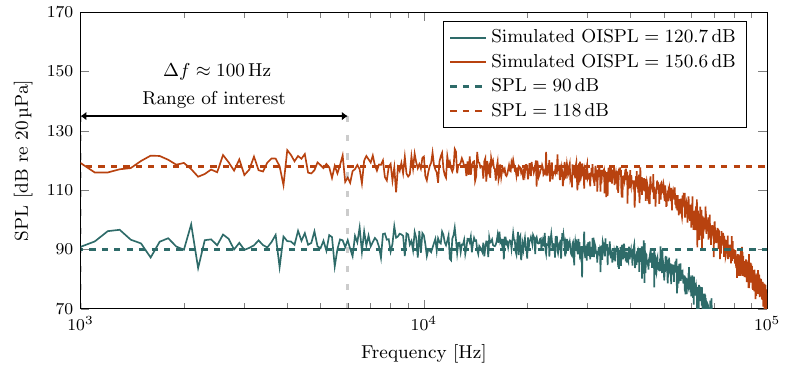}
    \caption{
        The simulated broadband sound source's power spectral density (PSD) follows the formulation of \citet{tam2010computational}.
        We follow \cref{eq:OISPL} to calculate the OISPL of the two plotted sources.
    }
    \label{Broadband_spec} 
\end{figure}

In addition to discrete tones, we also employed a broadband acoustic source, as described in \citet{tam2005computational, tam2010computational}, to investigate the energy dissipation of different acoustic sources in relation to leak effects.
The prescribed amplitude of the time-domain signal given in \eqref{eq4} is

\begin{gather}
    f_g(t) = \sum_{j=1}^{N} \sqrt{2 \Delta f_j S_j} \cos(f_j t + \chi_j)
    \label{eq4}
\end{gather} 
where $f_g(t)$ is the pressure amplitude as a function of time, $N$ is the total number of frequency bands in the prescribed spectrum, $\Delta f_j$ is the width of the $j$-th band of the prescribed spectrum, $S_j$ is the spectrum level at the $j$-th band, $f_j$ is the center frequency of the $j$-th band and $\chi_j$ is a random phase shift between $0$ and $\pi$ for the $j$-th band.
The prescribed bandwidth follows \cref{eq5}, which are
\begin{gather}
    \Delta f_j = r \, \Delta f_{j-1} 
    \quad \text{and} \quad
    \Delta f_1 = 10^4 \left( \frac{r-1}{r^N - 1} \right),
    \label{eq5}
\end{gather} 
where $r=1.01$, $\Delta f_1 = 500$, and $N = 100$.

MFC is GPU-accelerated and many-compute-node scalable so that we can conduct sufficiently large simulations in short wall-times~\citep{elwasif2023application,radhakrishnan22,radhakrishnan24,wilfong253}; further, it is permissively licensed and open-source.\footnote{Available at \url{https://github.com/MFlowCode/MFC}}
This performance advantage makes it useful for broadband acoustic noise simulations, which typically demand a few million time steps to determine the power absorption coefficients \cite{tam2005computational}.
The simulated PSD is shown in \cref{Broadband_spec}.
The overall incident sound pressure level (OISPL) is close to the prescribed value (\SI{120}{\decibel} and \SI{150}{\decibel}), and the average spectral level at the frequency range of interest (\SIrange[]{1}{6}{\kilo \hertz}) is plotted as the dashed lines.
Both \SI{120}{\decibel} and \SI{150}{\decibel} spectra are effectively flat, confirming that the source behaves as a broadband sound source.
This result shows the feasibility of using this method to simulate broadband acoustic waves with constant average SPL in the frequency range of interest.

\subsubsection{Simulation Configuration} \label{sss:sim_configuration}

\begin{figure}[htb]
    \centering
    \includegraphics[]{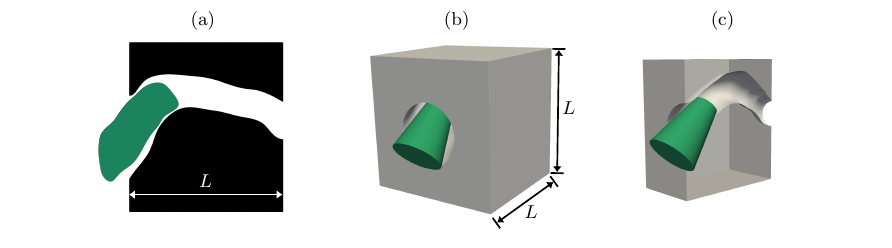}
    \caption{
        Model schematics (not to scale) of the (a) 2D ill-fitting earplug, (b) 3D ill-fitting earplug, and (c) half of the 3D ill-fitting earplug model to reveal the inner view of the model.
        The 3D earplug model is shaded in green for visual clarity.
    }
    \label{Sim_earplug_configuration} 
\end{figure}

Based on the methods implemented by \citet{wilfong2025mfc}, we used a ray-tracing algorithm to convert ASCII STL files, such as earplugs and ear canals, into structured computational grids as immersed boundaries \cite{tseng2003ghost} for applying the numerical methods in \cref{sss:sim_method}. 
The ear canal models are 2D and 3D (shown in \cref{Sim_earplug_configuration}) and follow the work of \citet{stinson1989specification}.
The distance between the inlet and outlet face of the 2D and 3D ear canal model is $L \approx \SI{25}{\milli \meter}$.
To the authors' knowledge, there has been no similar direct numerical study for realistic earplug leakage configurations.

We use uniformly structured meshes with equal cell sides.
We measured the system after multiple acoustic cycles to ensure it had reached a dynamic steady state.
The steady state occurs after all initial transients have vanished, and the acoustic periodic excitation repeats the same waveform every cycle.
We found that the fifth-order WENO scheme requires only $20$ cells per wavelength to resolve the open-field linearly propagating wave for the wavelength range studied in this work.
However, sound propagation in leakage is a wall-bounded problem, and the oscillatory boundary layer is a near-wall feature characterized by steep gradients, rather than a propagating wave.
Previous work has found that the acoustic-driven flow bounded by small slit resonators as small as $\SI{1}{\centi \meter}$ is laminar at a high incident level of approximately \SI{155}{\decibel}~\cite{tam2000microfluid}. 
The flow bounded by small air leaks is also laminar, even though vortical disturbance and its interaction with high-SPL sound can lead to the transition to turbulence outside the leaks.
The laminar condition with an oscillatory wave gives rise to a viscous Stokes layer inside small air leaks whose thickness is much smaller than the acoustic wavelength of our range of interest~\cite{tam2005computational, tam2010computational}.
Therefore, the Stokes layer resolution is the limiting factor in the grid resolution design, rather than the general acoustic wavelength.
According to \citet{white1991viscous} and \citet{tam2005computational}, the wavelength of the Stokes layer is $\lambda_s = 2 \sqrt{ {\pi\nu}/{f} }$, where $\nu$ is the kinematic viscosity and $f$ is frequency.

This study uses at least five cells per Stokes layer wavelength ($\lambda_s / \Delta x \ge 5$) for simulations to ensure negligible dissipation due to numerical errors.
The detailed simulation parameters are tabulated in \cref{table:sim_design}.
We conducted a grid convergence study and found that the power absorption coefficients with a doubled number of cells have negligible differences.

\begin{figure}[tb]
    \centering
    \includegraphics[]{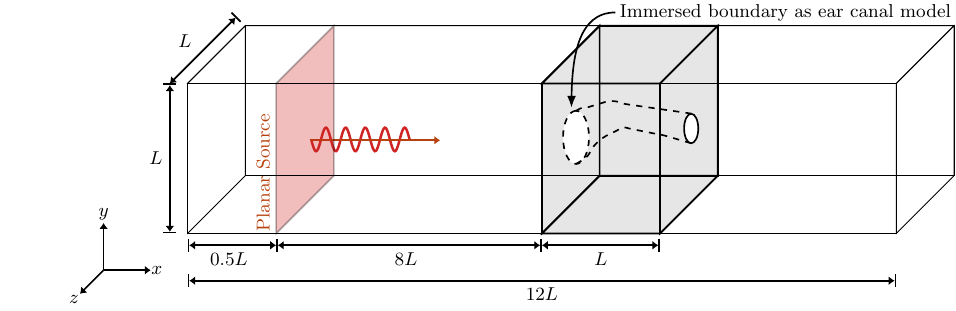}
    \caption{Schematic of the computational domain with acoustic source (not to scale) where the height and width of the domain ($L$) are much smaller than the wavelength of the sound in the interested frequency range.
    }
    \label{Sim_3Dearplug_schematics} 
\end{figure}

\Cref{Sim_3Dearplug_schematics} shows the 3D simulation domain schematics, where $L$ is the normal distance between the inlet and outlet of the ill-fitting earplug model.
The acoustic source plane is placed $L/2$ away from the left $x$-domain boundary, which performs ghost cell extrapolation and is non-reflective, just as the right $x$-domain boundary.
The direction of the sound wave is perpendicular to the $y$--$z$ plane.
An impermeable slip-wall boundary is used for the domain boundaries in the $y$ and $z$ directions.
The 2D simulation domain is similar to the $x$--$y$ slice of the 3D simulation domain, and the acoustic source plane behaves as a one-way line source in 2D.

{\small
\begin{table}
    \caption{
        Dimensionless simulation parameter specification.
        The grid resolution is uniform.
        The simulation duration is presented as the total number of time steps.
        \label{table:sim_design}
    }
    \centering
    \begin{tabular}{l c c r r}
    \toprule
    Case & Domain size  & Grid size & Time step size & Time steps \\
    \midrule
    2D Broadband & $12L \times L$ & $14400 \times 1200$ & $2.5\times 10^{-8} c_0/L$ & $5000000$ \\ 
    2D Discrete Tones & $12L \times L$ & $14400 \times 1200$ & $2.5\times 10^{-8} c_0/L$ & $1600000$ \\ 
    3D \SI{1}{\kilo\hertz} Discrete Tone & $12L \times L \times L$ & $ 3600 \times 300 \times 300$ & $1\times 10^{-7} c_0/L$ & $400000$  \\ 
    3D \SI{2}{\kilo\hertz} Discrete Tone & $12L \times L \times L$ & $5040 \times 420 \times 420$ & $7.5\times 10^{-8} c_0/L$& $540000$ \\ 
    3D \SI{3}{\kilo\hertz} Discrete Tone & $12L \times L \times L$ & $6240 \times 520 \times 520$ & $7.5\times 10^{-8} c_0/L$& $540000$ \\
    \bottomrule
    \end{tabular}
\end{table}
}

\subsubsection{2D Slit Resonator}
\label{sss:sim_validation}

\begin{figure}[htb]
    \centering
    \includegraphics[]{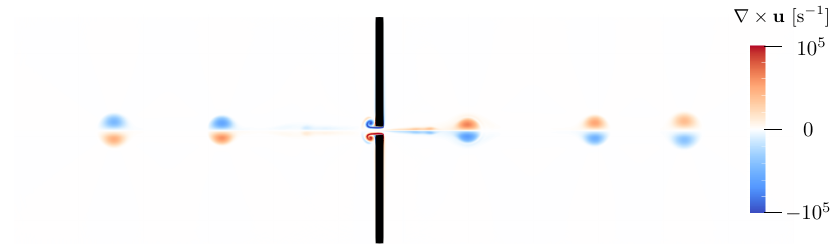}
    \caption{
        Vorticity distribution of a discrete tone propagating through the 2D slit resonator under dynamic steady conditions.  
    }
    \label{Sim_2DSlit_Visualization} 
\end{figure}

To verify the reliability of our numerical methods, we conducted a 2D simulation featuring a discrete tone ($\mathrm{ISPL} = \SI{150}{\decibel}$) and broadband noise ($\mathrm{OISPL} = \SI{150}{\decibel}$), modeled as a plane wave propagating through a 2D slit resonator.
The slit's opening and thickness ($d$) are scaled to be $1/28$ of the domain height to replicate the simulations performed in \citet{tam2001numerical}.
We use non-reflective ghost-cell extrapolation at the left horizontal boundary, and reflective impermeable slip walls for the remaining domain boundaries.
The simulation time and grid size were consistent with other 2D cases documented in \cref{table:sim_design}.
This framework facilitated the visualization of vortex shedding around the slit openings, a key feature depicted in \cref{Sim_2DSlit_Visualization}, which illustrates the resultant vortex shedding patterns.
This visual evidence supports our methodological approach to understanding wave propagation and vortex generation in high-intensity acoustic environments. 
This study also computes the power absorption coefficient for different frequencies.

\begin{figure}[htb]
    \centering
    \includegraphics[]{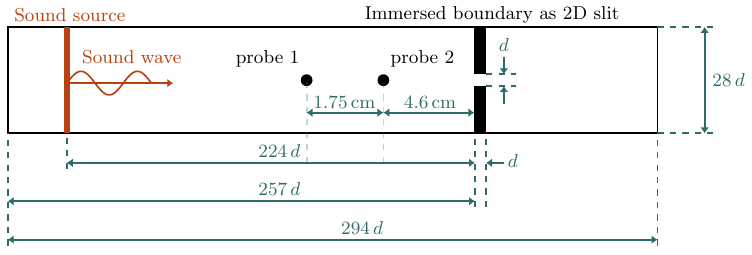}
    \caption{
        Schematic representation of the computational domain showing the acoustic source and the 2D slit (not to scale) to simulate the modeled ill-fitting earplug.
    }
    \label{Sim_validationschematics} 
\end{figure}

\begin{figure}[htb!]
    \centering
    \includegraphics[]{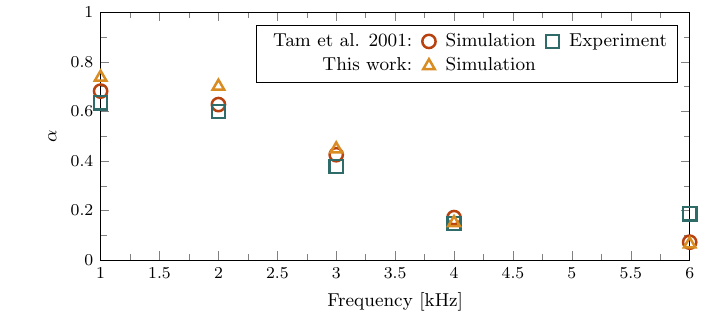}
    \caption{
        Comparison of power absorption coefficient spectra ($\alpha \approx 1- |\widehat{R}|^2 $ in this simulation) among numerical data and experiment of \citet{ahuja2000high} with discrete tones at $\mathrm{ISPL}= \SI{150}{\decibel}$.
    }
    \label{Sim_validationcompare} 
\end{figure}

For the 2D simulation described in \cref{Sim_validationschematics}, we recorded pressure at three points along the horizontal centerline of the domain. 
Two points are \SI{6.35}{\centi\meter} and \SI{4.6}{\centi\meter} upstream from the slit incident face, where higher-order modes are evanescent and lower-frequency aliasing errors are negligible.
Pressure was measured at each time step of the simulation, with the sampling frequency equal to the inverse of the simulation time step. 
The spectral analysis of the time history data followed the Welch's method \cite{welchsMethod}, yielding a frequency bandwidth of \SI{12.6}{\hertz}.

Following the methods of \citet{leung2007duct} and \citet{tam2010computational}, we employed the transfer function approach described by  \citet{chung1980transfer, chung1980transfer2} to decompose and calculate the incident and reflected acoustic powers.
The power reflection coefficient for discrete tones was determined via linear interpolation between two adjacent spectral points that bracketed each frequency. 
Given the reflective termination in our setup, the power absorption coefficient was assumed to be approximately $1- |\widehat{R}|^2$, where $|\widehat{R}|^2$ is the power reflection coefficient.
\Cref{Sim_validationcompare} compares the calculated absorption coefficient spectra and the results from previous studies, such as \citet{ahuja2000high}.
The results, obtained at a nominal incident sound intensity of \SI{150}{\decibel}, show alignment with established findings.
This agreement reinforces the accuracy and reliability of our numerical methods in simulating slit resonators under high-intensity acoustic conditions.

\section{Results}\label{s:results}

\subsection{Ill-fitting Earplug Experiments} \label{ss:exp_results}

Leak paths around actual earplugs may resemble those tested using orifice plates; however, critical differences in inlet geometry, orientation relative to incident noise, and material properties limit direct equivalence.
We measured the performance of both absorbing and non-absorbing earplugs using a modeled human ear canal, as described in \cref{TestArticles} of \cref{ss:exp_method}. 
To explore the maximum facility limit with respect to subsequent tests, we assessed the transmission loss (TL) of a solid 3D-printed plate (Test Article \#6) with the same thickness as the ear canal model.
Similar to the \SI{3.18}{\milli\metre} solid steel plate, the TL of the 3D-printed plate exceeds \SI{60}{\decibel}, indicating its performance is akin to that of a hard reflective wall. 

\begin{figure}[ht!]
    \centering
    \includegraphics[]{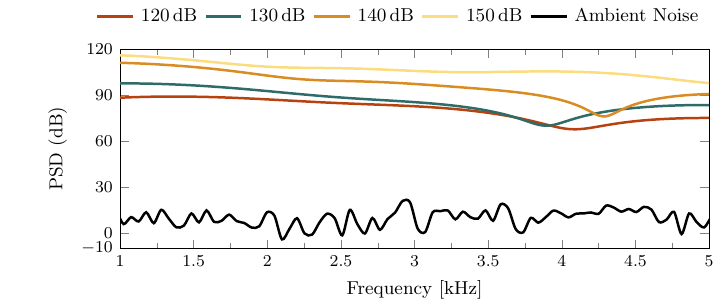}
    \caption{Power spectral density (PSD) of the ambient noise and incident impulsive signals with nominal OISPL of \SI{120}{\decibel}, \SI{130}{\decibel}, \SI{140}{\decibel}, and \SI{150}{\decibel}.
    }
    \label{Exp_incident} 
\end{figure}

Due to inherently low transmitted signals through well-sealed earplug configurations, the microphone on the transmission side often encountered signal-to-noise ratio (SNR) limitations when the overall incident sound pressure level (OISPL) was low. 
To optimize test conditions, we used pulses with OISPL exceeding \SI{120}{\decibel}, ensuring that all spectral data presented maintained an SNR of at least \SI{10}{\decibel}. 
\Cref{Exp_incident} shows the power spectral density (PSD) of the ambient noise and incident impulsive signals with nominal OISPL of \SIrange[]{120}{150}{\decibel}. 
The ambient noise maintains an average sound pressure level (SPL) of \SI{10}{\decibel} across the \SIrange[]{1}{5}{\kilo \hertz} range, while the minimum SPL measured for the lowest nominal OISPL condition (\SI{120}{\decibel}) exceeds \SI{65}{\decibel}.

\begin{figure}[ht]
    \centering
    \includegraphics[]{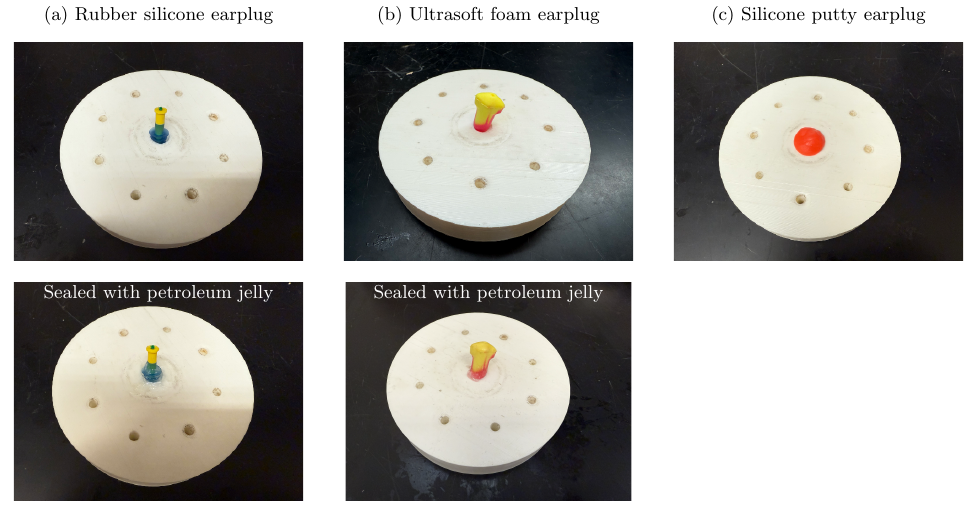}
    \caption{The photos of three types of commercial earplugs used on the ear canal model: (a) a silicone rubber earplug, (b) an ultrasoft foam earplug, and (c) a silicone putty earplug with five different configurations.
    }
    \label{Exp_Earplug_Configurations} 
\end{figure}

\begin{figure}[ht]
    \centering
    \includegraphics[]{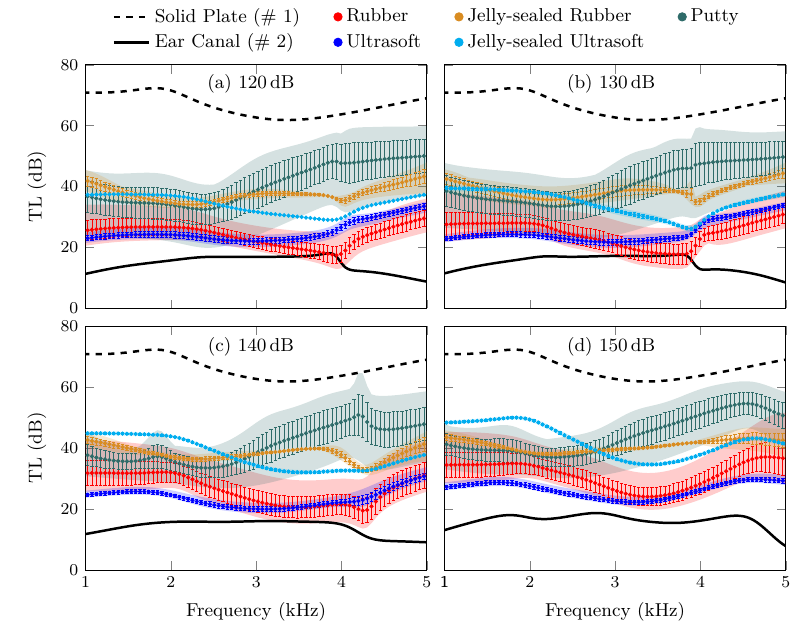}
    \caption{
        TL spectra for seven earplug–ear-canal configurations with nominal OISPL of (a) \SI{120}{\decibel}, (b) \SI{130}{\decibel}, (c) \SI{140}{\decibel}, and (d) \SI{150}{\decibel}.
        Circular markers (shown only for the five earplug designs) denote the arithmetic mean of ten independent insertions, the vertical bars indicate the standard deviation, and shaded regions span the full range observed across the ten trials.
        The 3D-printed plate and bare ear canal results remain unchanged because these configurations do not require manual insertion of earplugs; therefore, they appear without error bars.
    }
    \label{Exp_Earplug_Test_comparison_std} 
\end{figure}

The ear canal model, documented as Test Article \#7 in \cref{TestArticles}, is tested with three different types of earplugs, including absorbing ultrasoft foam, non-absorbing silicone putty, and non-absorbing silicone rubber, which fit loosely. 
Following the method described in \cref{ss:exp_method}, we convert the pressure time history measured at the two microphones into spectral levels. 
TL is defined in the same way as in \cref{ss:metrics} and is calculated according to \cref{eq:TL_tau} using the spectral levels. 
Five earplug configurations were studied: an ill-fitting silicone rubber earplug, an ultrasoft foam earplug with and without a petroleum jelly seal, and a well-fitting silicone putty earplug. 
\Cref{Exp_Earplug_Configurations} shows pictures of the five configurations. 
The silicone rubber and ultrasoft foam earplugs fit loosely into the ear canal, contrasting to the sealed fit achieved when petroleum jelly fully covers the visible leak paths around the earplugs.
The silicone putty, a hard modeling putty, is shaped to adequately cover the ear model openings, making it stiffer than the silicone rubber earplug and therefore capable of effectively sealing the ear canal model's inlet.

Each earplug was seated and tested ten times to characterize the variability introduced by the manual insertion of the earplugs.
The measured TL spectra of the five earplug configurations, along with the 3D-printed plates and the bare ear canal model, are plotted in \cref{Exp_Earplug_Test_comparison_std}, using incident pulses with varying OISPL (\SIrange[]{120}{150}{\decibel}). 
The spectral TL results show the arithmetic mean derived from ten independent insertion trials for each configuration shown in \cref{Exp_Earplug_Configurations}, with standard deviations represented by vertical bars and full ranges indicated by shaded regions. 
When a sealed fit was achieved, an average increase of \SI{18}{\decibel} TL was measured for the silicone rubber earplug and \SI{9}{\decibel} for the ultrasoft foam earplug across the frequency range under OISPL $= \SI{120}{\decibel}$. 
The TL increase for a sealed fit was reduced when the OISPL was increased. 
These experiments demonstrate that sealing leak paths substantially enhances earplug performance, highlighting the importance of fit and seal integrity.
Therefore, acoustic modeling of earplugs should prioritize accurately representing leak geometries and sealing conditions.

The comparable TL spectra of jelly-sealed silicone rubber and silicone putty suggest that the flexibility of the silicone putty provides a better seal than silicone rubber in practical applications.
Silicone rubber earplugs are unlikely to be fully inserted into the human ear to ensure a good fit, and they can be uncomfortable when fully inserted due to their high elasticity. 
Additionally, the TL standard deviation of non-absorbing earplugs (silicone rubber and putty) is larger than \SI{5}{\decibel}, whereas that of the absorbing earplug (i.e., ultrasoft foam) is less than \SI{2}{\decibel} across the \SIrange[]{1}{5}{\kilo \hertz} range. 
The TL results of the ill-fitting ultrasoft foam earplugs exhibit consistently smaller ranges and standard deviations compared to the ill-fitting silicone rubber earplugs, indicating less variability in manual insertion for ultrasoft foam earplugs.

\begin{figure}[ht]
    \centering
    \includegraphics[]{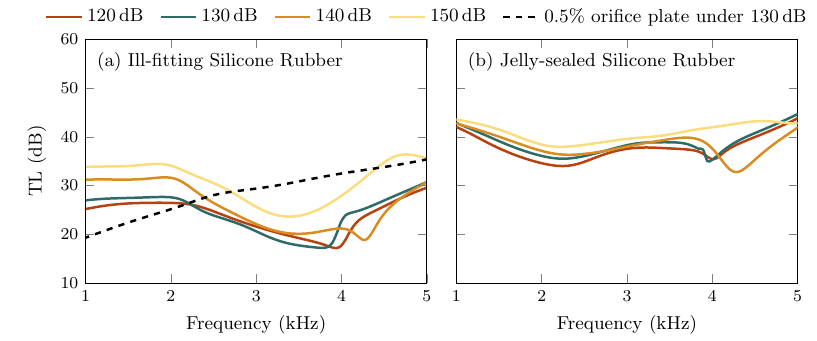}
    \caption{The transmission loss (TL) spectrum of (a) ill-fitting silicone rubber earplug compared to a 0.5\% orifice plate and (b) ill-fitting silicone rubber earplug sealed with petroleum jelly under an OISPL of \SIrange[]{120}{150}{\decibel}.
    }
    \label{Exp_Earplug_ill-fitting} 
\end{figure}

\Cref{Exp_Earplug_ill-fitting} shows the averaged TL spectra for the ill-fitting silicone rubber configuration tested with and without petroleum jelly sealing, evaluated across ten repetitions to assess the influence of the OISPL.
At frequencies below \SI{4}{\kilo\hertz}, the measured TL increases with rising OISPL, indicating a nonlinear or SPL-dependent attenuation mechanism.
At frequencies above \SI{4}{\kilo\hertz}, the TL values remain nearly constant within the tested OISPL range of \SIrange[]{120}{140}{\decibel}.

The PSD from \cref{Exp_incident} reveals a significant reduction in incident pulse energy around \SIrange[]{4}{4.5}{\kilo \hertz}, particularly for the \SI{140}{\decibel} case. 
This critical drop in incident energy likely contributes to the observed frequency-dependent behavior in the TL spectra from \cref{Exp_Earplug_ill-fitting}, as the incident energy no longer scales monotonically with increasing OISPL. 
Consequently, above \SI{4}{\kilo\hertz}, the limited incident energy available may reduce the manifestation of nonlinear effects that typically increase TL at higher SPLs.
Future experimental studies focusing on impulsive source modeling, with spectral analysis, are needed to explain the frequency-dependent behavior fully.

The TL of the ill-fitting silicone rubber case suggests that leak paths in an ill-fitting, non-absorbing earplug are comparable to those of a \SI{3.175}{\milli \meter}-long 0.5\% orifice, as described in \cref{sss:exp_orifice}.
Proper application of petroleum jelly around earplugs can enhance their performance by effectively sealing the leakage paths. 
The TL spectra of jelly-sealed silicone rubber suggest that effective sealing reduces nonlinear dissipation mechanisms, which appear at higher SPL. 
We need to measure orifice plates with much smaller openings to quantify the leaks in the jelly-sealed case using orifices. 
This setup requires machining orifices with diameters smaller than \SI{1}{\milli \meter} on steel plates or using TL tubes with larger inner diameters, where higher-order modes could propagate more easily between microphones.

\subsection{Two-dimensional Ill-fitting Earplug Simulation}
\label{ss:sim_2D_results}

\begin{figure}[ht]
    \centering
    \includegraphics[]{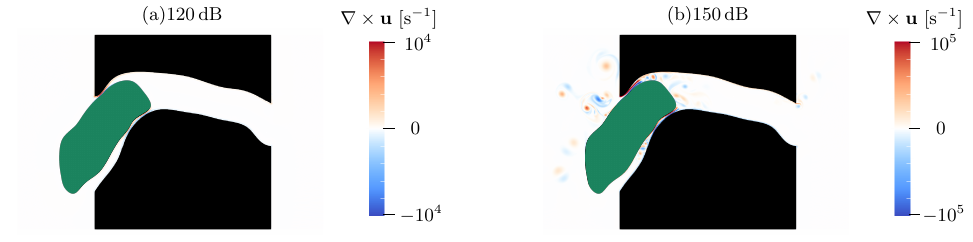}
    \caption{Instantaneous vorticity distribution of the simplified 2D earplug-canal geometry at the $30$-th cycle for \SI{1}{\kilo \hertz} signals at ISPL (a)~\SI{120}{\decibel} and (b)~\SI{150}{\decibel}.}
    \label{Sim_2Dearplug_omega3} 
\end{figure}

We simulated the 2D ear canal model with an ill-fitting earplug to observe the effects of varying ISPL. 
\Cref{Sim_2Dearplug_omega3} displays the vorticity distribution within the ear canal with a rigid ill-fitting earplug, evaluated at the $40$-th acoustic cycle to ensure a dynamic steady state at \SI{1}{\kilo\hertz} with \SI{120}{\decibel} and \SI{150}{\decibel}.
The color scale used for the \SI{150}{\decibel} scenario spans an order of magnitude greater than that for \SI{120}{\decibel}.

In the high ISPL scenario (\SI{150}{\decibel}), significant vorticity conversion is observed near sharp geometric discontinuities at both the ear canal inlet (around the leak path) and the outlet. 
These phenomena are minimal or absent under lower acoustic excitation (\SI{120}{\decibel}).
These findings confirm that higher acoustic pressures intensify velocity gradients near discontinuities, thereby increasing acoustic energy dissipation through the conversion of acoustic energy into vorticity.

\begin{figure}[htp!]
    \centering
    \includegraphics[]{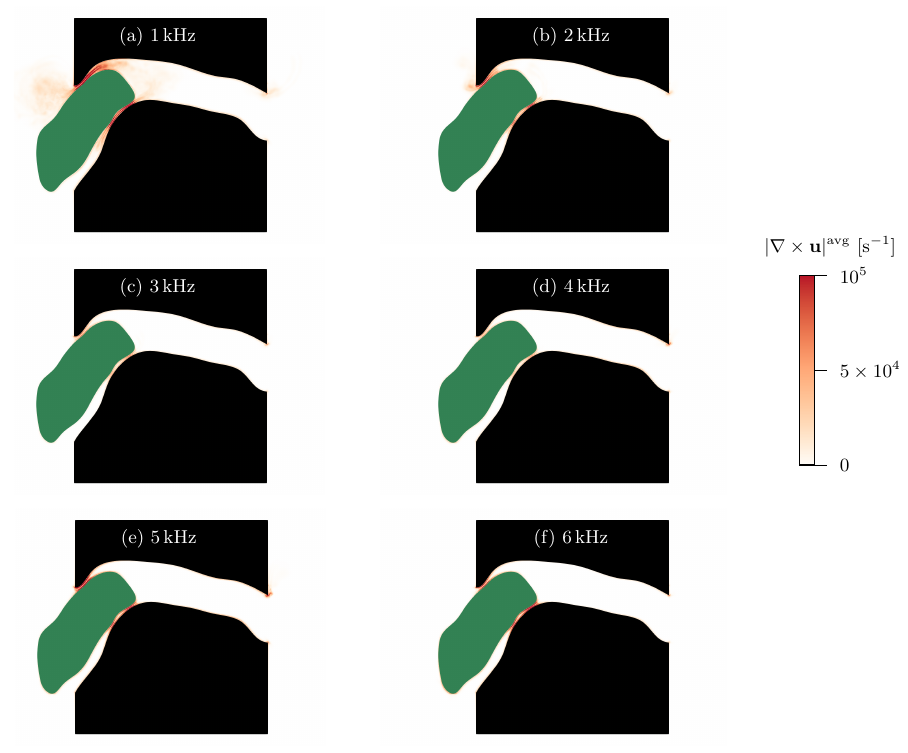}
    \caption{
    Time-averaged vorticity magnitude distributions in the simplified 2D earplug-canal geometry, subjected to an incident acoustic source at \SI{150}{\decibel} across six frequencies.
    The time averaging periods correspond to acoustic cycle ranges: (a) 10--40, (b) 20--80, (c) 30--120, (d) 40--160, (e) 50--200, and (f) 60--240. 
    Vortex shedding appears more predominant at lower frequencies.}
    \label{Sim_2Dearplug_vorticity} 
\end{figure}

\begin{figure}[htp!]
    \centering
    \includegraphics[]{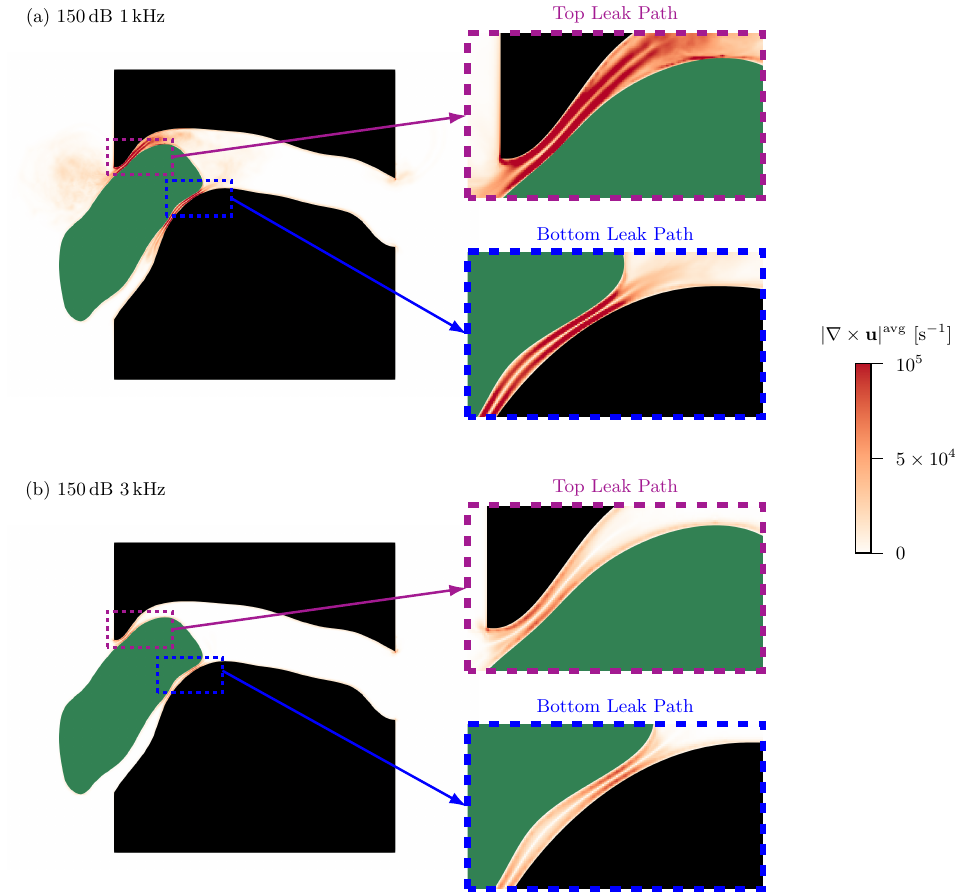}
    \caption{Magnified time-averaged vorticity magnitude distributions in the simplified 2D earplug-canal geometry, subjected to an incident acoustic source at \SI{150}{\decibel} for (a) \SI{1}{\kilo\hertz} and (b) \SI{3}{\kilo\hertz}.}
    \label{Sim_2Dearplug_vorticity_magnified} 
\end{figure}

\Cref{Sim_2Dearplug_vorticity} indicates that vorticity conversion in the leak path is strongly frequency-dependent. 
For example, when $\mathrm{ISPL}=\SI{150}{\decibel}$, the vorticity magnitude inside the leak paths at \SI{1}{\kilo \hertz} is notably larger than \SI{3}{\kilo \hertz}, as depicted in \cref{Sim_2Dearplug_vorticity_magnified}.
The time-averaged vorticity magnitude generally increases as the excitation frequency decreases, with an exception noted at \SI{5}{\kilo\hertz}, suggesting it may represent a resonant frequency of the ill-fitting earplug system.
We can explain the observed frequency-dependent trend through the Strouhal number ($\mathrm{St}$), defined as $\mathrm{St} = f L/a$, where $f$ is the acoustic frequency, $L$ is the characteristic length of the ill-fitting earplug system (the distance between the inlet and outlet of the ear canal), and $a$ is the speed of sound. 
This dimensionless parameter compares the acoustic oscillation period ($1/f$) with the convective timescale ($L/a$).

At lower frequencies, corresponding to $\mathrm{St} \ll 1$, the acoustic period is long enough for the fluid to flow through the narrow leak paths, develop into mature coherent vortices, and persist within the surrounding fluid. 
At higher frequencies, as $\mathrm{St}$ increases, the shorter acoustic periods restrict fluid particle motion. 
This results in rapid direction reversals that suppress vortex roll-up and inhibit the growth of stable vortical structures. 
Consequently, vorticity conversion is primarily confined to the immediate vicinity of the leak path.

These findings suggest that low-frequency acoustic excitation promotes greater attenuation through enhanced vortex-driven momentum exchange between the fluid inside and outside the leak path. 
This fluid-acoustic interaction mechanism effectively bypasses the acoustic isolation provided by the ill-fitting earplug at lower Strouhal numbers.

\begin{figure}[ht]
    \centering
    \includegraphics[]{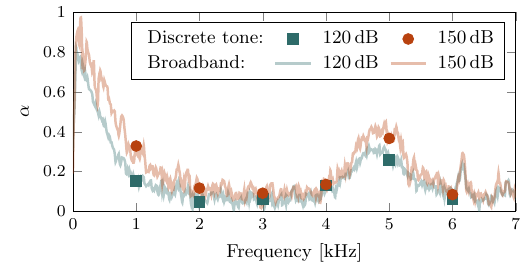}
    \caption{Comparison of the absorption coefficient spectra ($\alpha$ defined in \cref{ss:metrics}) of the modeled 2D ear canal with an ill-fitting earplug under an OISPL of \SI{120}{\decibel} and \SI{150}{\decibel}.}
    \label{Sim_2Dearplug_absoprtion} 
\end{figure}

We use the same metric and method described in the \cref{ss:metrics} to compute the power absorption coefficient based on the pressure time history at the numerical probes.
\Cref{Sim_2Dearplug_absoprtion} compares the absorption coefficient in different acoustic environments.
The simulated absorption coefficient of the larger ISPL is consistently higher than that of the smaller one across the presented frequency range.
At \SI{5}{\kilo \hertz}, where the maximum absorption is observed, the absorption coefficient of higher ISPL is approximately $0.1$ higher.
This result agrees with the time-averaged vorticity distribution and the experimental results shown in the \cref{ss:exp_results}. 
Higher absorption is measured at the ill-fitting earplug models in higher ISPL conditions.

\subsection{Three-dimensional Ill-fitting Earplug Simulation}
\label{ss:sim_3D_results}

Three-dimensional analyses are crucial as they account for realistic geometries and 3D flow-acoustic interactions. 
Using the same computational methods, we extended the high ISPL simulation to a 3D leakage study with a simple earplug model and the ear canal, as shown in \cref{sss:sim_configuration}. 
Similar to the 2D model, the geometry omits anatomical details such as the eardrum and middle-ear structures, allowing the analysis to focus solely on leakage-induced dynamics without the added complexity of termination effects.
The plane-wave assumption remains valid for the tested frequencies (\SIrange{1}{3}{\kilo\hertz}).

\begin{figure}[ht]
    \centering
    \includegraphics[]{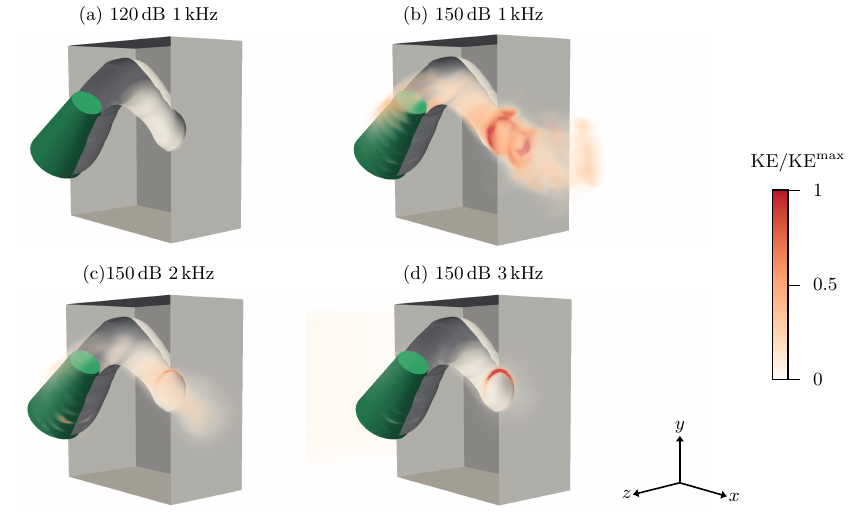}
    \caption{
        Normalized kinetic energy (KE) distributions at the $40$-th acoustic cycle in the simulated 3D ear canal model with a modeled ill-fitting earplug. 
    }
    \label{Sim_3Dearplug_3Dvis} 
\end{figure}

\begin{figure}[htp!]
    \centering
    \includegraphics[]{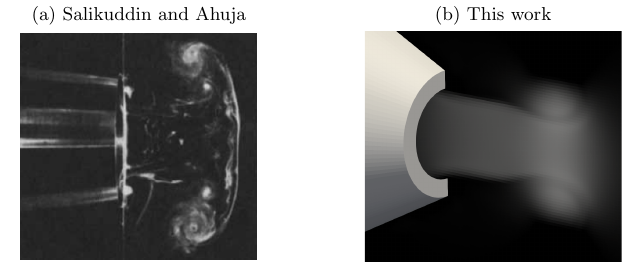}
    \caption{
        The vortex shedding at the outlet of our ear canal model is similar to the vortex shedding induced by a \SI{147}{\decibel} sound pulse at the outlet of a conical nozzle, as demonstrated in
        (a) experimental visualization using a smoke generator~\cite{salikuddin1983acoustic} and (b) numerical results from the present MFC simulation.}
    \label{Sim_nozzle} 
\end{figure}

\begin{figure}[ht]
    \centering
    \includegraphics[]{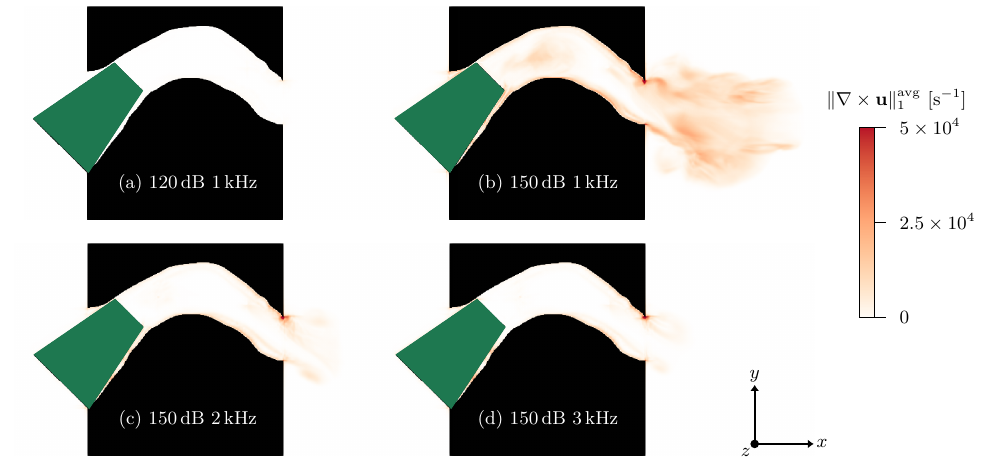}
    \caption{
        Time-averaged vorticity magnitude distributions in the 2D $x$--$y$ slice of the simulated 3D ear canal model with an ill-fitting earplug.
        The time-averaging periods correspond to acoustic cycle ranges: (a) and (b) 10--40, (c) 20--80,  and (d) 30--120.
        A larger vorticity magnitude and vortex shedding phenomena at lower frequencies indicate a greater potential for acoustic energy dissipation through conversion into vorticity.
    }
    \label{Sim_3Dearplug_vorticity}
\end{figure}

\begin{figure}[htp!]
    \centering
    \includegraphics[]{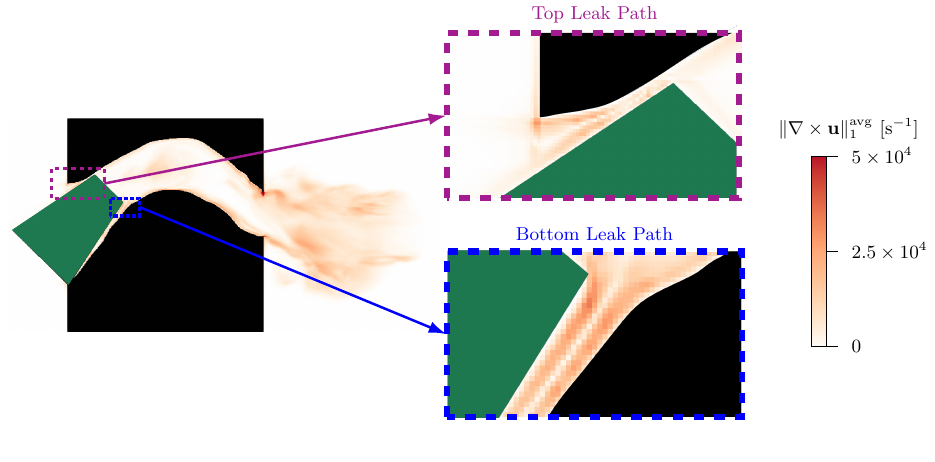}
    \caption{Magnified time-averaged vorticity magnitude distributions in the simulated 3D ear canal model with a modeled ill-fitting earplug, subjected to a \SI{1}{\kilo \hertz} incident acoustic source at \SI{150}{\decibel}.
    }
    \label{Sim_3Dearplug_vorticity_magnified} 
\end{figure}

\begin{figure}[ht!]
    \centering
    \includegraphics[]{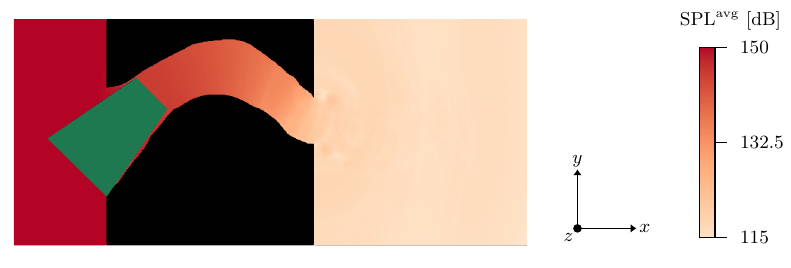}
    \caption{
    Time-averaged sound pressure level (SPL) distribution in the 2D $x$--$y$ slice of the simulated 3D ear canal model with an ill-fitting earplug, subjected to a \SI{1}{\kilo \hertz} incident acoustic source at \SI{150}{\decibel}.}
    \label{Sim_3Dearplug_SPL} 
\end{figure}

\Cref{Sim_3Dearplug_3Dvis} displays the instantaneous normalized kinetic energy (KE) distribution, evaluated at the $40$-th acoustic cycle, which ensures a dynamic steady state. 
KE values were normalized against the maximum obtained from the \SI{150}{\decibel} cases. 
At \SI{1}{\kilo\hertz}, high KE occurs near the ear canal exit, indicating acoustic energy dissipation through conversion into vorticity.
\Cref{Sim_nozzle} indicates the vortex shedding at the canal outlet is similar to experimentally visualized shear-layer roll-up produced by a sound pulse in a peak level of \SI{147}{\decibel} at the outlet of a conical nozzle with a diameter of \SI{6.2}{\centi \meter}~\cite{salikuddin1983acoustic}.
This comparison demonstrates that the outlet vortex shedding in the present model reflects a generic acoustic-driven termination behavior rather than a simulation artifact.
The time-averaged vorticity distributions in \cref{Sim_3Dearplug_vorticity}, taken in the middle $x$--$y$ plane, reveal the presence of vorticity near the exit of the ear canal model.
To show the detailed behavior around the leak path, \cref{Sim_3Dearplug_vorticity_magnified} shows a magnified view of the vorticity distribution for the \SI{150}{\decibel} \SI{1}{\kilo \hertz} test condition.
We observe that the sharp upper edge of the ear canal model exhibits a larger vorticity magnitude compared to the smoother lower edge at the canal outlet. 
This difference suggests that the sharp edge contributes to increased energy absorption, consistent with our 2D simulation results discussed in \cref{ss:sim_2D_results} and those by \citet{tam2005computational}.

Time-averaged vorticity magnitude distributions confirm increased dissipation near the ear canal exit region, with evident vortex shedding predominantly at lower frequencies, highlighting substantial conversion into vorticity.
Unlike the 2D simulation results in \cref{ss:sim_2D_results}, where primary vortex shedding occurs around the leak path, the 3D configuration shows the ear canal outlet as the primary site of vortex shedding.
In 3D, the complex geometry and spatial orientation of the leak path reduce the flow coherence required for sustained 3D vortex shedding, making the exit region a more favorable geometry for coherent vortex shedding.

\begin{figure}[ht!]
    \centering
    \includegraphics[]{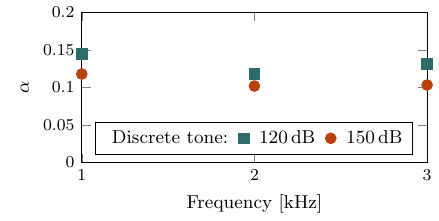}
    \caption{
        Comparison of the absorption coefficient spectra ($\alpha$ defined in \cref{ss:metrics}, \cref{eq:alpha}) of the modeled 3D ear canal and ill-fitting earplug for $\mathrm{ISPL}= \SI{120}{\decibel}$ and $\SI{150}{\decibel}$ as labeled.
        }
    \label{Sim_3Dearplug_absoprtion} 
\end{figure}

Additional confirmation is provided in the time-averaged SPL distribution in \cref{Sim_3Dearplug_SPL} for \SI{150}{\decibel} at \SI{1}{\kilo \hertz}, where we observed the SPL downstream of the leak path remains higher than \SI{125}{\decibel} and shows larger SPL gradients near the model outlet.
Together, these visualizations indicate that the ill-fitting earplug model still allows acoustic energy at unsafe SPL to leak through the modeled earplug and to be converted into vorticity, thus limiting their effectiveness in sound reduction and hearing protection.
Following the same metrics and methods described in \cref{ss:metrics,sss:sim_validation,Sim_3Dearplug_absoprtion}, we compare the power absorption coefficient under varying ISPLs.
As mentioned in \cref{sss:sim_configuration}, the power absorption coefficients have differences less than $5\%$ upon doubling the number of computational cells, indicating grid convergence.
We see a higher power absorption coefficient under higher ISPL conditions. 
This result aligns with our observations of vorticity distribution, confirming that conversion into vorticity dissipates acoustic energy more effectively under high ISPL conditions.

\section{Conclusion}\label{s:conclusions}

This study examined the TL of multiple commercial earplugs in a modeled ear canal and explored the effects of leakage around ill-fitting earplugs.
We combined an analytical model, experiments, and simulations under high SPL conditions. 
The results demonstrated the effectiveness of both experimental and computational methods, providing insights into the mechanisms of acoustic energy dissipation due to leakage.

Experimentally, we measured the TL of several earplug configurations using a custom-built two-sided transmission loss tube. 
Results revealed that leakage in ill-fitting earplugs substantially degrades earplug performance, reducing TL by an average of approximately \SI{18}{\decibel} for non-absorbing silicone rubber earplugs under an OISPL of \SI{120}{\decibel}.
Sealing leak paths with petroleum jelly greatly improved performance.
This result demonstrates a critical dependence on the seal's integrity for maintaining high attenuation, particularly in real-world applications where proper fitting might be challenging.

The ill-fitting ultrasoft foam earplugs exhibited narrower TL ranges and smaller standard deviations compared to silicone rubber and putty earplugs, indicating the sensitivity of silicone rubber and putty to variations in manual insertion. 
Additionally, we found that the TL of ill-fitting earplugs becomes increasingly sensitive to SPL, reflecting heightened dissipation mechanisms under high SPL conditions. 
At frequencies above \SI{4}{\kilo\hertz}, the limited incident energy of the impulsive signals prevented a monotonic increase of TL with OISPL.
This limitation highlights the importance of generating impulses with sufficient energy across a broader frequency range.
Future work should focus on developing a system capable of generating impulsive signals of varying widths and peak values to study the behavior of TL measurements for a wider frequency and SPL range.

Computationally, direct numerical simulations provided insight into vortex shedding within leak paths under high SPL conditions. 
Two-dimensional simulations revealed substantial conversion into vorticity at geometric discontinuities of the leak paths at high SPL.
Frequency-dependent behavior demonstrated greater vorticity generation and acoustic absorption at lower frequencies, which we explained by examining the Strouhal number relationship and its influence on vortex shedding. 
Three-dimensional simulations reinforced these observations and confirmed that realistic geometries have a significant impact on the effects of leakage and acoustic dissipation. 
The widespread vortices under high ISPL suggest that vorticity does not dissipate instantly at geometric discontinuities. 
This finding highlights the importance of future investigations into the mechanisms underlying energy conversion and dissipation among vortical structures.
Future work includes modeling an anatomically faithful ear canal model with middle-ear structures, such as the eardrum.

The main novelty of this study lies in integrating theoretical, experimental, and computational analyses to systematically quantify the mechanisms of acoustic leakage through ill-fitting earplugs under high SPL conditions.
The findings suggest practical strategies for improving earplug performance through careful attention to fit and seal integrity, particularly in high SPL environments, where leak-induced acoustic energy dissipation becomes critical.

\section*{Acknowledgments}

HY thanks Henry~Le~Berre for his guidance on the use of MFC and its features.
HY also thanks postdoctoral research fellow David~Nate~Ramsey in GTRI for the help in designing and building the experimental facility.
Associate Professor Julien Meaud, postdoctoral research fellow Tianyi Chu, and PhD student Zhixin Song from Georgia~Tech are thanked for helpful discussions.

\section*{Declaration of Conflicting Interests}

The authors declared no potential conflicts of interest with respect to the research, authorship, and/or publication of this article.

\section*{Funding}

This work was funded in part by the Georgia Institute of Technology Small Bets Internal Research and Development (IRAD) Program.
This work used Bridges2 at the Pittsburgh Supercomputing Center and NCSA Delta through allocation TG-PHY210084 (PI Spencer Bryngelson) from the Advanced Cyberinfrastructure Coordination Ecosystem: Services \& Support (ACCESS) program, which is supported by National Science Foundation grants \#2138259, \#2138286, \#2138307, \#2137603, and \#2138296.
SHB also acknowledges the resources of the Oak Ridge Leadership Computing Facility at the Oak Ridge National Laboratory, which is supported by the Office of Science of the U.S.\ Department of Energy under Contract No.\ DE-AC05-00OR22725.

\section*{Data Availability}

The data that support the findings of this study are available from the corresponding author upon reasonable request.
Codes are available at: \url{https://github.com/MFlowCode/MFC}.

\bibliography{main.bib}
\end{document}